\documentclass[divpdf,conference]{IEEEtran}
\usepackage{amssymb,stmaryrd,amsmath,amsfonts,rotating}
\usepackage{color}
\usepackage{bm}
\usepackage{latexsym}
\def\qed{\hfill $\Box$}

\usepackage{mathtools}
\mathtoolsset{showonlyrefs}

\newtheorem{corollary}{Corollary}
\newtheorem{theorem}{Theorem}
\newtheorem{discussion}{Discussion}
\newtheorem{definition}{Definition}

\newtheorem{lemma}{Lemma}

\newcommand{\Rpre}{R_\mathrm{pre}}

\newcommand{\dl}{d_{{l}}}

\newcommand{\dr}{d_{{r}}}
\newcommand{\dg}{d_{{g}}}

\allowdisplaybreaks
\usepackage{graphicx}	
\begin{document} 
\title{Spatially-Coupled Precoded  Rateless Codes}
\author{
Kosuke Sakata, Kenta Kasai and Kohichi Sakaniwa\\
\authorblockA{
Dept. of Communications and Integrated Systems, 
Tokyo Institute of Technology, 152-8550 Tokyo, Japan.\\
Email: {\tt \{sakata,kenta,sakaniwa\}@comm.ss.titech.ac.jp}} 
}

\maketitle
\begin{abstract}
Raptor codes are rateless codes that achieve the capacity on the binary erasure channels. 
However the maximum degree of optimal output degree distribution is unbounded. 
This leads to a computational complexity problem both at encoders and decoders. 
Aref and Urbanke investigated the potential advantage of universal achieving-capacity property of 
proposed spatially-coupled (SC) low-density generator matrix (LDGM) codes. 
However the decoding error probability of SC-LDGM codes is bounded away from 0. 
In this paper, we investigate  SC-LDGM codes concatenated with SC low-density parity-check codes.
The proposed codes can be regarded as SC Hsu-Anastasopoulos rateless codes.
We derive a lower bound of the asymptotic overhead from stability analysis for successful decoding by density evolution. 
The numerical calculation reveals that the lower bound is tight. 
We observe that with a sufficiently large number of information bits, the asymptotic overhead and 
the decoding error rate approach 0 with bounded maximum degree.
\end{abstract}
\section{Introduction}
Spatially-coupled (SC) low-density parity-check (LDPC) codes attract much attention due to 
their capacity-achieving performance under low-latency memory-efficient sliding-window belief propagation (BP) decoding.
The studies on SC-LDPC codes date back to the invention of convolutional LDPC codes by Felstr{\"o}m and Zigangirov \cite{zigangirov99}. 
Lentmaier {\itshape et al.}! observed that the BP threshold of regular SC-LDPC codes coincides with the maximum a posterior (MAP) threshold of the underlying block LDPC 
codes with a lot of accuracy  by density evolution  \cite{lentmaier_II}. 
Kudekar {\it et al.}~ proved that SC-LDPC codes achieve the MAP threshold of BEC \cite{5695130} 
and the binary-input memoryless output-symmetric (BMS) channels \cite{2012arXiv1201.2999K} under BP decoding. 

Rateless codes are a class of erasure-recovering codes which produce limitless sequence of encoded bits 
from $k$ information bits so that receivers can recover the $k$ information bits from arbitrary $(1+\alpha)k/(1-\epsilon)$ received symbols from BEC($\epsilon$). 
We denote  {\itshape overhead} by $\alpha$. 
Designing rateless codes with vanishing overhead is desirable, which implies the codes achieve the capacity of BEC($\epsilon$). 
LT codes \cite{1181950} and  raptor codes \cite{raptor} are rateless codes that
achieve vanishing overhead $\alpha\to 0$ in the limit of large information size over the BEC.
By a nice analogy between the BEC and the packet erasure channel \textcolor{black}{(e.g., Internet)}, 
rateless codes have been successfully adopted by several industry standards.

A raptor code can be viewed as concatenation of an outer high-rate LDPC code and infinitely many single parity-check codes of length $d$, 
where $d$ is chosen randomly with probability $\Omega_d$ for $d\ge 1$. 
Raptor codes need to have unbounded maximum degree $d$ for $\Omega_d\neq 0$. 
This leads to a computation complexity problem both at encoders and decoders. 

The authors presented empirical results in \cite{HSU_MN_IEICE}  showing 
that SC MacKay-Neal (MN) codes and SC Hsu-Anastasopoulos (HA) codes achieve the capacity of BEC with bounded maximum degree. 
Recently a proof for SC-MN codes are given in \cite{ISIT_OJKP}. 
It was observed that the SC-MN codes and SC-HA codes have the BP threshold close to the Shannon limit in \cite{6400949} over BMS channels.

Aref and Urbanke \cite{ITW_AREF_URBANKE} investigated the potential advantage of universal achieving-capacity property of 
SC low-density generator matrix (LDGM) codes. 
They observed that the decoding error probability steeply decreases with overhead $\alpha=0$ with bounded maximum degree over various BMS channels. 
However the decoding error probability was proved to be bounded away from 0 with bounded maximum degree for any $\alpha$. 
This is explained from the fact that there are a constant fraction of bit nodes of degree 0. 

In this paper, we investigate  SC-LDGM codes concatenated with SC-LDPC codes. 
The proposed codes can be regarded as SC-HA rateless codes.
We derive a lower bound of the asymptotic overhead from stability analysis for successful decoding by density evolution. 
The numerical calculation reveals that the lower bound is tight. 
We observe that with a sufficiently large number of information bits, the asymptotic overhead and the decoding error rate approach 0 with bounded maximum degree.

\section{Encoder and Decoder}
\subsection{Encoder}
Let $k$ denote the number of information bits. 
We define a $(\dl,\dr,\dg,L,w)$ code for $\dl\ge 2,\dr\ge 2,\dg\ge 2$ as follows. 
The $(\dl,\dr,\dg,L,w)$ code are defined on $L$ sections from 0 to $L-1$. 
Each section has $M$ pre-coded bits. 
Note that, in \cite{5695130}, $2L+1$ sections  $[-L,+L]$ were considered. 
Instead, for the sake of simplicity, we consider $L$ sections in $[0,L-1]$. 
First, the $k$ information bits are pre-coded with $(\dl,\dr,L,w)$  codes \cite{5695130} into $LM$ bits 
$x({0,0}),\dotsc,x({L-1,M-1})$. 
In this paper,  we assume that the bits in the $i$-th section for $i\in [0,L-1]$ are transmitted and the bits in other sections are shortened. 
Namely, the shortened bits are set to 0 and are not transmitted. 
Let $\Rpre(L)$ denote the design coding rate of $(\dl,\dr,w,L)$ codes. In \cite{5695130}, $\Rpre(L)$ is given by
\begin{align}
 \Rpre(L)
&=1-\frac{\dl}{\dr}- \frac{\dl}{\dr}\frac{w-1-2\sum_{i=1}^{w-1}(i/w)^{\dr}}{L}\label{Rpre}\\
&\stackrel{L\to\infty}{=}1-\frac{\dl}{\dr}.
\end{align}
It follows that $k=   \Rpre(L)LM$. 

After encoding the $k$ bits into $LM$ coded bits by pre-code, the $LM$ pre-coded bits further will be  encoded by  an inner code as follows. 
Repeat the following procedure endlessly for $t\in [1,\infty)$.
\begin{enumerate}
 \item Choose a section $i^{(t)}\in[0,L+w-2]$ uniformly at random from $L+w-1$ sections.
 \item Choose $\dg$ section shifts  $j^{(t)}_1,\dotsc,j^{(t)}_{\dg}\in [0,w-1]$ with repetition uniformly at random.
 \item Choose $\dg$ bit-indices $l^{(t)}_1,\dotsc,l^{(t)}_{\dg}\in[0,M-1]$ with repetition  uniformly at random.
 \item Add $\dg$ bits and transmit the sum as
	   \begin{align}
		x({i^{(t)}-j^{(t)}_1,l^{(t)}_1})+\cdots+x({i^{(t)}-j^{(t)}_{\dg},l^{(t)}_{\dg}}).\label{154801_11Jan13}
	   \end{align}
\end{enumerate}

\subsection{Decoder}
Assume that transmission takes place over BEC($\epsilon$) and we have  $n$ received symbols $y^{(1)},\dotsc,y^{(n)}$ each of which is  0, 1 or `?'. 
Define the {\itshape overhead} $\alpha$ as 
\begin{align}
 \alpha=\frac{n}{k}(1-\epsilon)-1. \label{201403_16Aug12}
\end{align}
In this setting,  we have $(1+\alpha)k=n(1-\epsilon)$ unerased received symbols. 
Independence of the coding scheme ensures that we can assume, without loss of generality, that time indices of $n$ received  symbols are arbitrary. 
For simplicity, we assume that the receiver receives $n$ symbols at time $t=1,\dotsc,n$ without loss of generality. 

We assume that the decoder knows $i^{(t)}$, $\dg$ section shifts $j^{(t)}_1,\dotsc,j^{(t)}_{\dg}$ and 
bit-indices $l^{(t)}_1,\dotsc,l^{(t)}_{\dg}$ in \eqref{154801_11Jan13} for each received symbol at time $t=1,\dotsc,n$.
From these information and the knowledge of the precode, one can construct a factor graph  for sum-product decoding \cite{910572}. 
The factor graph consists of $LM$ variable nodes (bit nodes) $x({0,0}),\dotsc,x({L-1,M-1})$
and $(1-\Rpre(L))LM$ parity-check factor nodes (check nodes) of pre-code and factor nodes (channel nodes) of factor
\begin{align}
  \mathbf{1}[x({i^{(t)}-j^{(t)}_1,j^{(t)}_1})+\cdots+x({i^{(t)}-j^{(t)}_{\dg},j^{(t)}_{\dg}})=y^{(t)}]
\label{161033_11Jan13}
\end{align}
for $t=1,\dotsc,n$, where $\mathbf{1}[\hspace{0.4mm}\cdot\hspace{0.4mm}]$ is defined as 1 if the argument is true and 0 otherwise. 
We say that the factor node of factor \eqref{161033_11Jan13} is in the section $i^{(t)}$. 
\section{Performance Analysis}
In this section, we investigate the performance of the coupled rateless codes and derive a bound. 
\subsection{Performance Analysis by Density Evolution}
In this subseciton, we derive the density evolution update equation. 
The following lemma clarifies the degree distributions of inner codes. 
\begin{lemma}
Let $\Lambda_d$ be the probability that a bit node has $d$ neighboring channel nodes.
Let $\beta$ be the average number of channel nodes adjacent to a bit node.
In the limit of large $M$, we have
\begin{align}
 &\beta=\frac{\dg}{1-\epsilon}\frac{L\Rpre(L)(1+\alpha)}{L+w-1}, \label{beta}\\
 &\sum_{d\ge 0}\Lambda_dx^d =e^{-\beta(1-x)}=\sum_{d\geq0}\frac{{\beta}^de^{-{\beta}}}{d!}x^d.
\end{align}
\end{lemma}
{\itshape Proof}: Let $N$ denote the average number of channel nodes per section. 
There are $L+w-1$ sections containing channel nodes. 
We have $n$ channel nodes in total.
\begin{align}
 N&=\frac{n}{L+w-1}=\frac{1}{1-\epsilon}\frac{(1+\alpha)k}{L+w-1}\\
 &=\frac{1}{1-\epsilon}\frac{ (1+\alpha)\Rpre(L)LM}{L+w-1}, 
\end{align}
where we used $k=   \Rpre(L)LM$. 
Recalling that $\beta$ is the average number of channel nodes adjacent to a bit node, we have 
 \begin{align}
\beta&=\frac{\dg N}{M}. 
 \end{align}
Equation \eqref{beta} immediately follows from this. 
Each section has  $N$ channel nodes of degree $\dg$, in other words, we have $\dg N$ edges in each section. 
Let $\Lambda_d$ denote the probability that a bit node in the $i$-th section has $d$ channel nodes within sections from $i$ to $i+w-1$. 
Since  each channel node is generated independently, 
the probability $\Lambda_d$ follows  a binomial distribution as follows. 
\begin{equation*}
\label{L_d}
\Lambda_d={\dg N\choose d}\left(\frac{1}{ M}\right)^d\left(1-\frac{1}{ M}\right)^{\dg N-d}
\end{equation*}
The probability generating function of $\Lambda_d$ is given as follows. 
 \begin{align}
 \Lambda(x)&:=\sum_{d\geq 0}\Lambda_dx^d=\left(\frac{x}{ M}+1-\frac{1}{ M}\right)^{\dg N} \\
 &\stackrel{M \to \infty}{=}\exp[-{\beta} (1-x)]=\sum_{d\geq0}\frac{{\beta}^de^{-{\beta}}}{d!}x^d. 
 \end{align}
This implies  $\Lambda_d=\frac{{\beta}^de^{-{\beta}}}{d!}$ in the limit of $M \to \infty$. In other words, the degree $d$ follows the 
Poisson distribution of average ${\beta}$. 
\qed

Let us describe density evolution update equations. 
Let $p_i^{(\ell)}$ and $s_i^{(\ell)}$ be the erasure probability of messages sent from bit nodes in the $i$-th section to check nodes and channel nodes, respectively,  at the $\ell$-th iteration of BP decoding
of $(\dl,\dr,\dg,L,w)$ codes in the limit of large $M$. 
The density evolution \cite{910577} gives update equations for $p_i^{(\ell)}$ and $s_i^{(\ell)}$ as follows. 
For $i\notin [0,L-1]$,  $p_i^{(\ell)}=s_i^{(\ell)}=0$.
For $i\in [0,L-1]$, $p_i^{(0)}=s_i^{(0)}=1$, and for $\ell\ge 0$, 
\begin{align}
  p_i^{(\ell+1)}&=\Bigl(\frac{1}{w}\sum_{j=0}^{w-1}\bigl(1-(1-\frac{1}{w}\sum_{k=0}^{w-1}p_{i+j-k}^{(\ell)})^{d_r-1}\bigr)\Bigr)^{d_l-1}\\
&\cdot\Lambda\Bigl(\frac{1}{w}\sum_{j=0}^{w-1}\bigl(1-(1-\epsilon)(1-\frac{1}{w}\sum_{k=0}^{w-1}s_{i+j-k}^{(\ell)})^{d_g-1}\bigr)\Bigr), \\
   s_i^{(\ell+1)}&=\Bigl(\frac{1}{w}\sum_{j=0}^{w-1}\bigl(1-(1-\frac{1}{w}\sum_{k=0}^{w-1}p_{i+j-k}^{(\ell)})^{d_r-1}\bigr)\Bigr)^{d_l}\\
&\cdot\lambda\Bigl(\frac{1}{w}\sum_{j=0}^{w-1}\bigl(1-(1-\epsilon)(1-\frac{1}{w}\sum_{k=0}^{w-1}s_{i+j-k}^{(\ell)})^{d_g-1}\bigr)\Bigr), 
\end{align} 
where $\lambda(x)=\frac{\Lambda'(x)}{\Lambda'(1)}=\exp[-{\beta} (1-x)]=\Lambda(x)$.

Let $\mathbb{P}_\mathrm{b}^{(\ell)}$ be the decoding error probability at the $\ell$-th iteration of BP decoding given as follows.
\begin{align}
 \mathbb{P}_\mathrm{b}^{(\ell)}&:=\frac{1}{L}\sum_{i=1}^{L}p_i^{(\ell)}.
\end{align}
\begin{definition}\label{dif}
One can easily check $\mathbb{P}_\mathrm{b}^{(\ell)}$ has its limit $\mathbb{P}_\mathrm{b}^{(\infty)}(L):=\lim_{\ell\to\infty}\mathbb{P}_\mathrm{b}^{(\ell)}(L)$
since $\mathbb{P}_\mathrm{b}^{(\ell)}$ is 
decreasing in $\ell$. 
We define {\itshape overhead threshold} $\alpha^{\ast}_{L}$ and its corresponding $\beta^{\ast}_L$ as follows.  
\begin{align}
 &\alpha^{\ast}_{L}:=\inf\bigl\{\alpha>0\mid \mathbb{P}_\mathrm{b}^{(\infty)}({L})=0\bigr\},\\
 &\beta^{\ast}_{L}:=\inf\bigl\{\beta>0\mid \mathbb{P}_\mathrm{b}^{(\infty)}({L})=0\bigr\}.
\end{align}
We say  $(\dl,\dr,\dg,L,w)$ codes {\itshape achieve the capacity of} BEC($\epsilon$) if 
\begin{align}
 \limsup_{L\to\infty}\alpha^{\ast}_{L}=0.
\end{align}
\end{definition}
\begin{discussion}
We will explain why we exclude the case $\dg=1$.
Assume  $\dg=1$. The density evolution update equations can be  reduced as follows. 
\begin{align}
  p_i^{(\ell+1)}&=
\begin{cases}
\displaystyle \Lambda(\epsilon)\Bigl(\frac{1}{w}\sum_{j=0}^{w-1}\bigl(1-(1-\frac{1}{w}\sum_{k=0}^{w-1}p_{i+j-k}^{(\ell)})^{d_r-1}\bigr)\Bigr)^{d_l-1}\\
\hfill(i\in[0,L-1]),\\
0\hfill (i\notin[0,L-1]).
\end{cases}
\end{align} 
This is equivalent to the density evolution update equation of the precode that is a $(\dl,\dr,w,L)$ code transmitted over BEC($\Lambda(\epsilon)$) \cite{5695130}. 
If the error probability goes to 0, $\Lambda(\epsilon)$ has to be less than the Shannon limit $\Lambda(\epsilon) = e^{-\beta_L^{\ast}(1-\epsilon)} < 1-\Rpre(L)$.
It follows that $\beta_L^{\ast}$ is bounded as follows. 
 \begin{align}
 \beta_L^{\ast} > \frac{1}{1-\epsilon} \ln \frac{1}{1-\Rpre(L)}. 
 \end{align}
From  \eqref{betaLW} we have
 \begin{align}
  \alpha_L^\ast &> \frac{L+w-1}{L\Rpre(L)} \ln \frac{1}{1-\Rpre(L)}-1\\
  &\stackrel{L\to\infty}{=} \frac{\dr}{\dr-\dl}\ln\frac{\dr}{\dl} -1>0. 
 \end{align}
This implies the $(\dl,\dr,\dg=1,L,w)$ codes do not achieve the capacity of BEC($\epsilon$). This is the reason why we exclude the case $\dg=1$ in this paper. 
\end{discussion}
\begin{lemma}\label{limbeta}
 The $(\dl,\dr,\dg,L,w)$ codes achieve the capacity of BEC($\epsilon$) if and only if 
 \begin{align}
\limsup_{L\to\infty}\beta^\ast_{L}&=\frac{\dg}{1-\epsilon}\Bigl(1-\frac{\dl}{\dr}\Bigr). \label{betast}
 \end{align}
\end{lemma}
{\itshape Proof}:
This is straightforward from \eqref{beta}, we have 
\begin{align}
 \beta^{\ast}_{L}&=\frac{\dg}{1-\epsilon} \Rpre(L)\frac{L}{L+w-1}(1+\alpha^{\ast}_{L}) \label{betaLW}\\
&=\frac{\dg}{1-\epsilon}\Bigl(1-\frac{\dl}{\dr}\Bigr) \quad (L\to\infty). 
\end{align}
\qed
\subsection{Performance Bound by Stability Analysis}
In the following theorem, we  derive a lower  bound of overhead threshold $\alpha^\ast_L$. 
\begin{theorem}\label{P_beta}
For $(\dl=2,\dr,\dg,L,w)$ codes, 
if  $\mathbb{P}_\mathrm{b}^{(\infty)}({L})=0 $
then there exist $\underline{\alpha}^{\ast}_{L}$ and $\underline{\beta}^{\ast}_{L}$ such that 
\begin{align}
& \alpha^{\ast}_{L}\ge\underline{\alpha}^{\ast}_{L},\quad \beta^{\ast}_{L}\ge \underline{\beta}^{\ast}_{L}, \label{Phituyou}\\
 & 
\lim_{L\to\infty}\underline{\alpha}^{\ast}_{L}= \max\Bigl[\frac{{\ln(\dr-1)}}{\dg(1-2/\dr)}-1,0\Bigr],\label{225002_6Jan13}\\
 & 
\lim_{L\to\infty}\underline{\beta}^{\ast}_{L}=\max\Bigl[\frac{\ln(\dr-1)}{1-\epsilon},\frac{\dg}{1-\epsilon}\Bigl(1-\frac{\dl}{\dr}\Bigr)\Bigr]. 
\end{align}
\end{theorem}

{\itshape Proof:}
Let $P_{L}$ denote an $L\times L$ matrix whose $(i,j)$ entry is $\frac{\partial p_{i}^{(\ell+1)} }{\partial p_{j}^{(\ell)}}$.
As we will see, this does not depend on $\ell$.
Let $\rho(P_{L})$ denote the spectral radius  of $P_{L}$.
We will derive a lower bound of $\rho(P_{L})$. 

Some calculation reveals that at $\bm{p}^{(\ell)}=\bm{s}^{(\ell)}=\bm{0}$
for $\dl = 2$. 
\begin{align}
\frac{\partial p_{i}^{(\ell+1)} }{\partial p_{j}^{(\ell)}}
&=\frac{(\dr-1)\lambda(\epsilon)}{w^2}\frac{\partial}{\partial p_{j}^{(\ell)}}\sum_{l=0}^{w-1}\sum_{k=0}^{w-1}p_{i+l-k}^{(\ell)}\\
&=\left\{ \begin{array}{ll}
				      \frac{w-|i-j|}{w^{2}}(\dr-1)\lambda(\epsilon) & (|i-j|\leq w) \\
					      0 & (|i-j| > w)
       \end{array} \right.\label{dl=2}
      \end{align}
and $\frac{\partial p_{i}^{(\ell+1)} }{\partial p_{j}^{(\ell)}}=0$ for $\dl > 2$.
It holds that for $\dl\ge 2$,
\begin{align}
\frac{\partial p_{i}^{(\ell+1)} }{\partial s_{j}^{(\ell)}}=
\frac{\partial s_{i}^{(\ell+1)} }{\partial p_{j}^{(\ell)}}=
\frac{\partial s_{i}^{(\ell+1)} }{\partial s_{j}^{(\ell)}}=0. 
\end{align}
at $\bm{p}^{(\ell)}=\bm{s}^{(\ell)}=\bm{0}$.
We drop $\ell$ since \eqref{dl=2} is independent of $\ell$.

From \eqref{dl=2}, we can see that $P_{L}$ is a {\itshape positive band matrix of width} $w$,  which is defined in Definition \ref{def:band} in Appendix. 
Since $P_{L}$ is a positive band matrix of width $w$, one can see that $P_{L}$ is irreducible from Lemma \ref{obi} in Appendix.
Let $\lambda_1,\dotsc,\lambda_L$ be the eigenvalues of $P_{L}$, recall that $\rho(P_{L})$ is the spectral radius of $P_L$.
We have
\begin{align}
& \rho(P_{L}) := \max_i(|\lambda_i|). 
\end{align}
Since $P_{L}$ is symmetric, the eigenvalues are real. 

Let $\lambda_1>\dotsc>\lambda_L$ be the eigenvalues of $P_{L}$. 
Perron-Frobenius theorem \cite{1406483} asserts that the eigenvalue that gives the spectral radius of a non-negative irreducible matrix is positive.
Since $P_{L}$ is non-negative symmetric irreducible matrix, the eigenvalue that gives spectral radius of $P_{L}$ is positive.
Then we have \begin{align}
 \rho(P_{L}) = \lambda_1. \label{lam}
\end{align}

For $\delta>0$, we define $\beta:=\beta^{\ast}_{L}+\delta$.
Since $\beta>\beta^{\ast}_{L}$, it follows $\mathbb{P}_\mathrm{b}^{(\infty)}({L})=0$.
From \eqref{lam},  we have for $\forall\bm{x} \in \mathbb{R}^{L} \setminus \{\bm{0}\}$, 
\begin{align}
1&>\rho( P_{L} )\stackrel{(a)}{=} \max_{\bm{x} \in \mathbb{R}^{L}:\bm{x}\neq \bm{0}}\frac{\bm{x}^{\mathsf{T}} P_L \bm{x} } {\bm{x}^{\mathsf{T}}\bm{x}}
\ge\frac{\bm{1}^{\mathsf{T}} P_L \bm{1} } {\bm{1}^{\mathsf{T}}\bm{1}} \\ 
&=  (\dr-1)e^{-\beta(1-\epsilon)}\frac{w^2L- (w-1)w(w+1)/3 } {w^{2}L}\\
 &\stackrel{L\to\infty}{=}(\dr-1)e^{-\beta^{\ast}_{L}(1-\epsilon)}, \label{165611_25Jan13}
\end{align}
where we used  \cite[Theorem 4.2.2]{0521386322} for (a). 
Solving $\beta$ from this inequality, we obtain
\begin{align}
 \beta >
&\frac{1}{1-\epsilon}\ln\Bigl[(\dr-1)\bigl(1-\frac{(w-1)(w+1) } {3wL}\bigr)\Bigr].\label{201242_13Jan13}
\end{align}
$\lim_{\delta\to 0}\beta=\beta^{\ast}_{L}$ denote that $\beta^\ast_{L}\ge \mbox{RHS of } \eqref{201242_13Jan13}$.
A trivial lower bound $\alpha_L^{\ast} \ge 0$ is true, since we can not surpass the capacity. 
From this and \eqref{betaLW}, it follows that 
\begin{align}
{\beta}^{\ast}_{L}&\ge\max\Bigl[\mbox{RHS of } \eqref{201242_13Jan13}, \frac{\dg}{1-\epsilon} \Rpre(L)\frac{L}{L+w-1} \Bigr]=:\underline{\beta}^{\ast}_{L}\\
\alpha^{\ast}_{L}
&\ge
\frac{ \underline{\beta}^{\ast}_{L}({1-\epsilon})({L+w-1})}{{\dg}{L} \Rpre(L)}-1
=:\underline{\alpha}^{\ast}_{L}	
\end{align}
In the limit of large $L$, we have
\begin{align}
\lim_{L\to\infty}\underline{\beta}^{\ast}_{L} &=\max\Bigl[\frac{\ln(\dr-1)}{1-\epsilon},\frac{\dg}{1-\epsilon}\Bigl(1-\frac{\dl}{\dr}\Bigr)\Bigr],\\
 \lim_{L\to\infty}\underline{\alpha}^{\ast}_{L}&= \max\Bigl[\frac{{\dr\ln(\dr-1)}}{\dg(\dr-2)}-1,0\Bigr].
\end{align}
This concludes Theorem \ref{P_beta}. 
\begin{discussion}
For $L\ge 2w-1$, $P_{L}$ have entries taking value from 1 to $w$. 
From \cite[Lemma 5.6.10]{0521386322}, we can bound $\rho(P_L)$ as follows. 
\begin{align}
\rho(P_L)&\le \|P_{L}\|_1:= \max_{1\leq i \leq L} \sum_{j=1}^{L} |(P_{L})_{i,j}|\\
&= (\dr-1)e^{-\beta(1-\epsilon)}\frac{1}{w^2}\Bigl(w+2\sum_{i=1}^{w-1}i\Bigr)\\
&= (\dr-1)e^{-\beta(1-\epsilon)}\label{norm}
\end{align}
From this, we can see that the bound \eqref{165611_25Jan13} is tight for large $L$. 
\end{discussion}
\qed

\begin{corollary}\label{CA}
For capacity-achieving $(\dl=2,\dr,\dg,L,w)$  codes have to satisfy
 \begin{align}
 \dg \ge \frac{\dr \ln (\dr-1)}{\dr-2}.\label{125116_21Jan13} 
 \end{align}
 This condition is not satisfied for $\dr=2$ or $\dg=2$.
\end{corollary}
{\itshape Proof:} From Definition \ref{dif}, capacity-achieving codes satisfy $\underline{\alpha}^{\ast}_{L}$ goes to 0 in the limit of large $L$. 
To be precise, 
\begin{align*}
\lim_{L\to\infty}\underline{\alpha}^{\ast}_{L}= \max\Bigl[\frac{{\dr\ln(\dr-1)}}{\dg(\dr-2)}-1, 0\Bigr] = 0. 
\end{align*}
The inequality \eqref{125116_21Jan13} immediately follows from this. 
\qed

\section{Decoding Performance}
In this section, we demonstrate the decoding performance of the $(\dl,\dr,\dg,L,w)$ codes.

\begin{figure}[t]
\begin{picture}(200,170)(0,0)
\put(0,0){\includegraphics[width=0.48\textwidth]{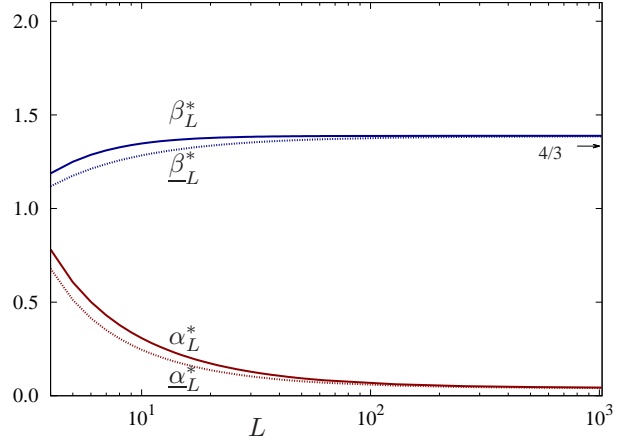} }
\put(100,00){$L$}
\put(70,35){$\alpha^{\ast}_{L}$}
\put(210,105){\scriptsize{4/3}}
\put(70,20){$\underline{\alpha}^{\ast}_{L}$}
\put(70,120){$\beta^{\ast}_{L}$}
\put(70,100){$\underline{\beta}^{\ast}_{L}$}
\end{picture}
\caption{
The asymptotic overhead $\alpha^{\ast}_L$ and the average degree of $\beta^{\ast}_L$ and their lower bounds $\underline{\alpha}^{\ast}_L$ and $\underline{\beta}^{\ast}_L$ of $(\dl=2,\dr=3,\dg=2,L,w=2)$ codes over BEC($\epsilon$=0.5). 
The asymptotic overhead threshold  $\alpha^{\ast}_L$ does not converge to 0 since the codes do not satisfy the condition of Corollary \ref{CA}. 
Figure suggests the lower bounds are tight for large $L$.
}
\label{fig:alpha}
\end{figure}

Figure \ref{fig:alpha} shows convergence 
the overhead threshold $\alpha^{\ast}_L$ and  $\beta^{\ast}_L$ and their lower bounds $\underline{\alpha}^{\ast}_L$ and $\underline{\beta}^{\ast}_L$ of $(\dl=2,\dr=3,\dg=2,L,w=2)$ codes over BEC($\epsilon$=0.5). 
The codes do not satisfy the condition of Corollary \ref{CA}. 
This explains why  $\alpha^{\ast}_L$ does not converge to 0 and $\beta^{\ast}_L$ does not converge to 4/3 which is given in Lemma \ref{limbeta} 
as the limiting value of capacity-achieving codes.  
We observe that 
 $\alpha^{\ast}_L$ approaches $\underline{\alpha}^{\ast}_{\infty}=\frac{3{\ln(2)}-2}{2}\simeq 0.03972$ and
 $\beta^{\ast}_L$ approaches $\underline{\beta}^{\ast}_{\infty}=2\ln(2)\simeq  1.38629$ 
which suggest the lower bounds are tight for large $L$.

\begin{figure}[t]
\begin{picture}(200,170)(0,0)
\put(0,0){\includegraphics[width=0.48\textwidth]{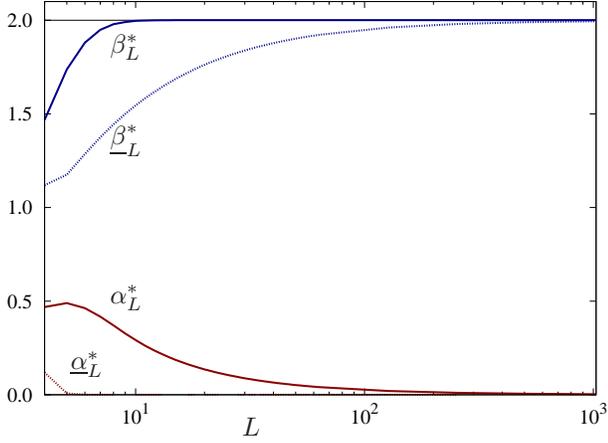} }
\put(100,00){$L$}
\put(50,50){$\alpha^{\ast}_{L}$}
\put(35,25){$\underline{\alpha}^{\ast}_{L}$}
\put(50,145){$\beta^{\ast}_{L}$}
\put(50,110){$\underline{\beta}^{\ast}_{L}$}
\end{picture}
\caption{
The asymptotic overhead $\alpha^{\ast}_L$ and the average degree of $\beta^{\ast}_L$ and their lower bounds $\underline{\alpha}^{\ast}_L$ and $\underline{\beta}^{\ast}_L$ of $(\dl=2,\dr=3,\dg=3,L,w=2)$ codes over BEC($\epsilon$=0.5). 
The codes  satisfy the condition of Corollary  \ref{CA}. 
We observe that  $\alpha^{\ast}_L$ approaches $\underline{\alpha}^{\ast}_{\infty}=0$ and
 $\beta^{\ast}_L$ approaches $\underline{\beta}^{\ast}_{\infty}=2$, 
which suggest the lower bounds are tight for large $L$.
}
\label{fig:alpha2}
\end{figure}

Figure \ref{fig:alpha2} shows convergence 
the asymptotic overhead threshold $\alpha^{\ast}_L$ and the average degree of $\beta^{\ast}_L$ and their lower bounds $\underline{\alpha}^{\ast}_L$ and $\underline{\beta}^{\ast}_L$ of $(\dl=2,\dr=3,\dg=3,L,w=2)$ codes over BEC($\epsilon$=0.5). 
The codes  satisfy the condition of Corollary  \ref{CA}. 
Thought this does not necessarily ensure $\alpha^{\ast}_L$ approaches  0 and $\beta^{\ast}_L$ approaches 2 which is given in Lemma \ref{limbeta} as the limiting value of capacity-achieving codes, this is likely the case. 
We observe that 
 $\alpha^{\ast}_L$ approaches $\underline{\alpha}^{\ast}_{\infty}=0$ and
 $\beta^{\ast}_L$ approaches $\underline{\beta}^{\ast}_{\infty}=2$, 
which suggest the lower bounds are tight for large $L$.

\begin{figure}[t]
\begin{picture}(200,170)(0,0)
\put(0,70){\rotatebox{90}{$\alpha^{\ast}_L$}}
\put(0,0){\includegraphics[width=0.48\textwidth]{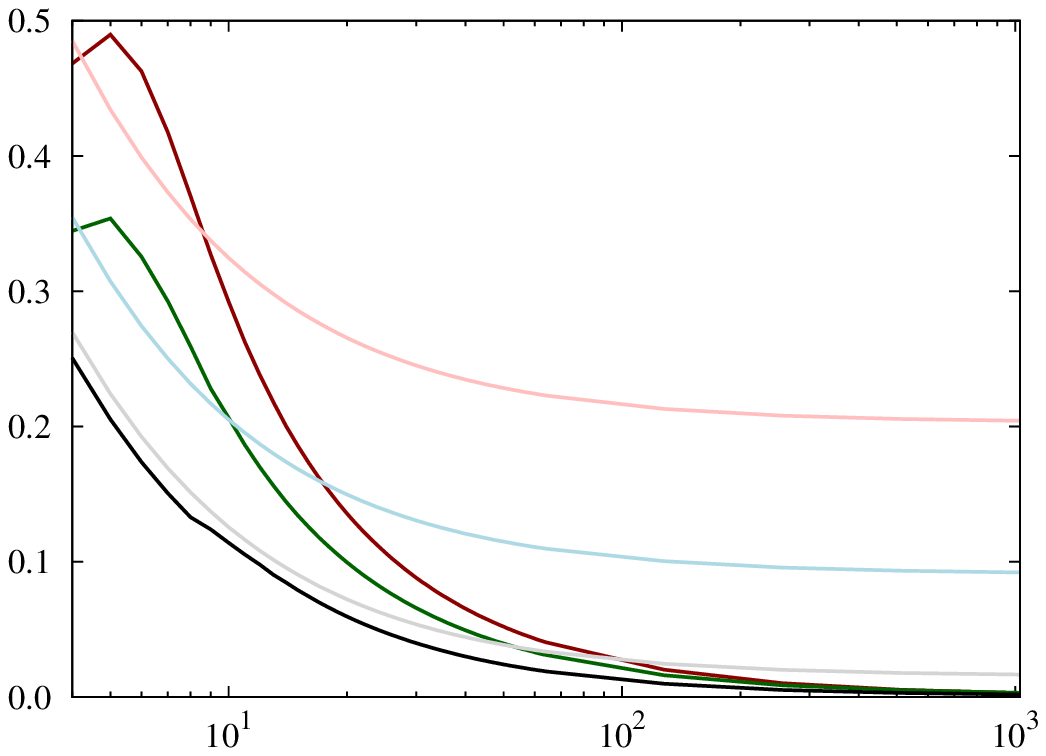} }
\put(45,150){\rotatebox{-70}{\scriptsize{$\dr=3$}}}
\put(30,130){\rotatebox{-50}{\scriptsize{$\dr=4$}}}
\put(30,70){\rotatebox{-50}{\scriptsize{$\dr=14$}}}
\put(180,25){\rotatebox{0}{\scriptsize{$\dr=15$}}}
\put(180,80){\rotatebox{0}{\scriptsize{$\dr=30$}}}
\put(180,45){\rotatebox{0}{\scriptsize{$\dr=20$}}}
\put(100,00){$L$}
\end{picture}
\caption{
Comparison of approaching speed of overhead threshold $\alpha^{\ast}_L$ of $(\dl=2,\dr, \dg=3,L,w=2)$ codes with $\dr\in \{3,4,14,15,20,30\}$ over BEC($\epsilon$=0.5).
The codes with  $\dr \le 14 $ satisfy the condition of Corollary \ref{CA}, while  the codes with  $\dr > 14$ do not. 
The fastest approaching speed is attained at  $\dr=14$. 
}
\label{fig:speed}
\end{figure}

Figure \ref{fig:speed} compares approaching speed of overhead threshold $\alpha^{\ast}_L$ of $(\dl=2,\dr, \dg=3,L,w=2)$ codes with $\dr\in \{3,4,14,15,20,30\}$ over BEC($\epsilon$=0.5).
The codes of  $\dr \le 14 $ satisfy the condition of Corollary \ref{CA}, while the codes of  $\dr > 14$ do not.
The fastest approaching speed is attained at  $\dr=14$. 

\section{Conclusion}
We propose spatially-coupled precoded regular rateless codes.
We have derived a lower bound $\underline{\alpha}^{\ast}_L$ of asymptotic overheads threshold $\alpha^{\ast}_L$. 
The numerical calculation of density evolution shows that the bound is tight for large coupling number $L$ and 
asymptotic overheads threshold $\alpha^{\ast}_L$ goes to 0 for large $L$ with bounded density. 
The possible future work is an extension to BMS channels and a proof for capacity-achievability. 
\section*{Acknowledgements}
The second author would like to thank V.~Aref for helping and discussing  this work. 
The second author started this work with V.~Aref when he stayed at EPFL in 2011. 
\bibliographystyle{IEEEtran} 
\bibliography{IEEEabrv,../../kenta_bib}

\begin{thebibliography}{10}
\providecommand{\url}[1]{#1}
\csname url@rmstyle\endcsname
\providecommand{\newblock}{\relax}
\providecommand{\bibinfo}[2]{#2}
\providecommand\BIBentrySTDinterwordspacing{\spaceskip=0pt\relax}
\providecommand\BIBentryALTinterwordstretchfactor{4}
\providecommand\BIBentryALTinterwordspacing{\spaceskip=\fontdimen2\font plus
\BIBentryALTinterwordstretchfactor\fontdimen3\font minus
  \fontdimen4\font\relax}
\providecommand\BIBforeignlanguage[2]{{%
\expandafter\ifx\csname l@#1\endcsname\relax
\typeout{** WARNING: IEEEtran.bst: No hyphenation pattern has been}%
\typeout{** loaded for the language `#1'. Using the pattern for}%
\typeout{** the default language instead.}%
\else
\language=\csname l@#1\endcsname
\fi
#2}}

\bibitem{zigangirov99}
A.~J. Felstr{\"o}m and K.~S. Zigangirov, ``Time-varying periodic convolutional
  codes with low-density parity-check matrix,'' \emph{{IEEE} Trans. Inf.
  Theory}, vol.~45, no.~6, pp. 2181--2191, June 1999.

\bibitem{lentmaier_II}
M.~Lentmaier, D.~V. Truhachev, and K.~S. Zigangirov, ``To the theory of
  low-density convolutional codes. {II},'' \emph{Probl.~ Inf.~ Transm.~},
  no.~4, pp. 288--306, 2001.

\bibitem{5695130}
S.~Kudekar, T.~Richardson, and R.~Urbanke, ``Threshold saturation via spatial
  coupling: Why convolutional {LDPC} ensembles perform so well over the
  {BEC},'' \emph{{IEEE} Trans. Inf. Theory}, vol.~57, no.~2, pp. 803--834, Feb.
  2011.

\bibitem{2012arXiv1201.2999K}
S.~{Kudekar}, T.~{Richardson}, and R.~{Urbanke}, ``{Spatially Coupled Ensembles
  Universally Achieve Capacity under Belief Propagation},'' \emph{ArXiv
  e-prints}, Jan. 2012.

\bibitem{1181950}
M.~Luby, ``{LT} codes,'' in \emph{Proc. 40th Annual Allerton Conf. on Commun.,
  Control and Computing}, 2002, pp. 271 -- 280.

\bibitem{raptor}
A.~Shokrollahi, ``Raptor codes,'' \emph{{IEEE} Trans. Inf. Theory}, vol.~52,
  no.~6, pp. 2551--2567, June 2006.

\bibitem{HSU_MN_IEICE}
K.~Kasai and K.~Sakaniwa, ``Spatially-coupled {M}ac{K}ay-{N}eal codes and
  {H}su-{A}nastasopoulos codes,'' \emph{IEICE Trans. Fundamentals}, vol. E94-A,
  no.~11, pp. 2161--2168, Nov. 2011.

\bibitem{ISIT_OJKP}
N.~Obata, Y.-Y. Jian, K.~Kasai, and H.~D. Pfister, ``Spatially-coupled
  multi-edge type ldpc codes with bounded degrees that achieve capacity on the
  bec under bp decoding,'' Jan. 2013, submitted to ISIT2013.

\bibitem{6400949}
D.~G.~M. Mitchell, K.~Kasai, M.~Lentmaier, and D.~J. Costello, ``Asymptotic
  analysis of spatially coupled {M}ac{K}ay-{N}eal and {H}su-{A}nastasopoulos
  {LDPC} codes,'' in \emph{2012 International Symposium on Information Theory
  and its Applications (ISITA)}, Oct. 2012, pp. 337--341.

\bibitem{ITW_AREF_URBANKE}
V.~Aref and R.~Urbanke, ``Universal rateless codes from coupled {LT} codes,''
  in \emph{Proc. 2011 {IEEE} Information Thoery Workshop (ITW)}, Oct. 2011.

\bibitem{910572}
F.~Kschischang, B.~Frey, and H.-A. Loeliger, ``Factor graphs and the
  sum-product algorithm,'' \emph{{IEEE} Trans. Inf. Theory}, vol.~47, no.~2,
  pp. 498--519, Feb. 2001.

\bibitem{910577}
T.~Richardson and R.~Urbanke, ``The capacity of low-density parity-check codes
  under message-passing decoding,'' \emph{{IEEE} Trans. Inf. Theory}, vol.~47,
  no.~2, pp. 599--618, Feb. 2001.

\bibitem{1406483}
S.~Pillai, T.~Suel, and S.~Cha, ``The {P}erron-{F}robenius theorem: some of its
  applications,'' \emph{IEEE Signal Processing Magazine}, vol.~22, no.~2, pp.
  62 -- 75, Mar. 2005.

\bibitem{0521386322}
R.~A. Horn and C.~R. Johnson, \emph{Matrix Analysis}.\hskip 1em plus 0.5em
  minus 0.4em\relax Cambridge University Press, 1990.

\end{thebibliography}
\appendix
\begin{definition}
A square matrix $A$  is said to be a reducible matrix when there exists a permutation matrix $P$ such that
$P^{\mathsf{T}}AP= 
\begin{pmatrix}
 X&Y\\
0&Z
\end{pmatrix},
$ where X and Z are both square. Otherwise $A$ is said to be irreducible.

\end{definition}
\begin{definition}
  Let $A$ be a square matrix of size $m$. 
 The graph $\mathcal{G}(A)$ of $A$ is defined to be the directed graph on $m$ nodes ${N_1,...,N_m}$ 
 in which there is a directed edge leading from $N_i$ to $N_j$ if and only if $a_{i,j}\neq 0$.
 $\mathcal{G}(A)$ is called strongly connected if for each pair of nodes $(N_i, N_j)$ there is a sequence of directed edges leading from $N_i$ to $N_j$.
\end{definition}
The following lemma can be found in \cite[p.~362]{0521386322}. 
\begin{lemma}
\label{irr}
 A square matrix  $A$ is an irreducible matrix if and only if $\mathcal{G}(A)$ is strongly connected.
\end{lemma}

\begin{definition}
\label{def:band}
 We say that a square real matrix $A=(a_{i,j})$ is positive band matrix of width $w$ if
 \begin{align}
 a_{i,i+j}
 \left\{ \begin{array}{ll}
				      > 0 & (|j| \leq w) \\
					 =     0 & (|j| > w)\\
      \end{array} \right.
 \end{align}
\end{definition}
\begin{lemma}\label{obi}
 $L\times L$ matrix $A_{L}$ is irreducible if $A_{L}$ is  positive band  matrix of width $w\ge 1$. 
\end{lemma}
{\itshape Proof:} 
From Definition  \ref{def:band}, it holds that for any $0\le j\le L$, $N_j\in\mathcal{G}(A)$ is adjacent to $N_{\max(0,{j-1})}$ and $N_{\min({j+1},L-1)}$. 
The lemma follows readily from Lemma \ref{irr}. 
\qed

\end{document}